# Extending Sparse Tensor Accelerators to Support Multiple Compression Formats

Eric Qin*, Geonhwa Jeong*, William Won*, Sheng-Chun Kao*, Hyoukjun Kwon*,
Sudarshan Srinivasan†, Dipankar Das†, Gordon E. Moon‡, Sivasankaran Rajamanickam§ and Tushar Krishna*
**Georgia Tech*, Atlanta, USA. {ecqin,geonhwa.jeong,william.won,felix,hyoukjun}@gatech.edu, tushar@ece.gatech.edu
†*Intel Labs*, Bangalore, India. {sudarshan.srinivasan, dipankar.das}@intel.com
‡*Korea Aerospace University*, Goyang, Republic of Korea. ehmoon@kau.ac.kr
§*Sandia National Laboratories*, Albuquerque, USA. srajama@sandia.gov

***Abstract*—Sparsity, which occurs in both scientific applications and Deep Learning (DL) models, has been a key target of optimization within recent ASIC accelerators due to the potential memory and compute savings. These applications use data stored in a variety of compression formats. We demonstrate that both the compactness of different compression formats and the compute efficiency of the algorithms enabled by them vary across tensor dimensions and amount of sparsity. Since DL and scientific workloads span across all sparsity regions, there can be numerous format combinations for optimizing memory and compute efficiency. Unfortunately, many proposed accelerators operate on one or two fixed format combinations. This work proposes hardware extensions to accelerators for supporting numerous format combinations seamlessly and demonstrates ∼4× speedup over performing format conversions in software.**

## I. Introduction

Scientific computing and deep learning (DL) applications are impacting society in many ways, from solving computational chemistry, power network, robotics and economics problems [1] to generating personalized recommendations, image classifications, and machine translations [2]. Unfortunately, the compute kernels and tensors that enable these applications can become extremely large (Amazon Review tensor has dimension sizes in millions [3]), leading to high memory and compute costs. This has led to growing interest in domain-specific accelerators for tensor algebra [4], [5]. In addition to custom compute units for tensor operations, the key property these accelerators exploit is *sparsity* in tensors within DL and scientific applications. For instance, modern Deep Neural Networks (DNNs) often exhibit 30% to 90% sparsity in trained weights without loss in accuracy via state-of-the-art pruning techniques [6], [7], while tensors in scientific workloads are often more than 99% sparse [5]. This naturally leads to the question on the right compression format for storing the tensors and there has been growing research on this topic [8], [9], [10], [11], [12] (see Sec. II).

The two main criteria for choosing a specific compression format depend on both (1) *compactness*, which represents the total memory size required to store the metadata and data for the tensor, and (2) *compute efficiency*, which represents the access complexity and data-to-compute mappings. High compactness is critical for storage, as a data transfer from DRAM can cost 6400× more energy than an add operation [13], [14]. High compute efficiency is critical to maintain high utilization within the accelerator[1], which in turn translates to runtime and energy-efficiency. To separate these two objectives, we refer to the format used for storage in memory as Memory Compression Format (MCF) and the format used during the computation of the algorithm within the accelerator as Algorithm Compression Format (ACF). If the MCF and ACF are different, naturally a converter is needed. Fig. 1a presents an overview of this concept with three matrices: **A**, **B**, and **O**.

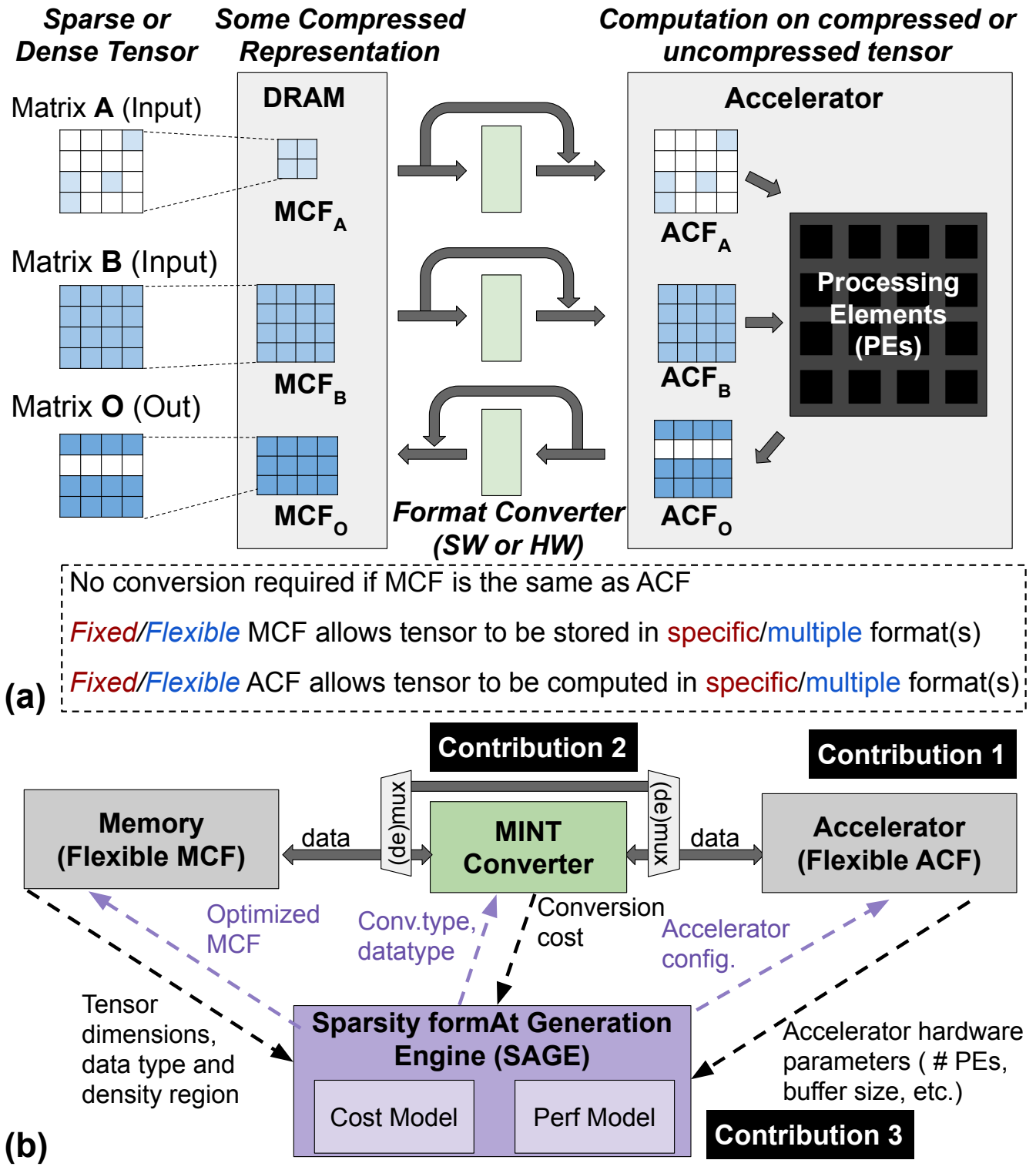

Fig. 1: (a) Memory Compression Format (MCF) and Algorithm Compression Format (ACF). (b) Our research contributions: (1) accelerator extensions for flexible ACFs, (2) MINT (format converter), and (3) SAGE (format recommender).

There have been limited studies evaluating the trade-offs of various MCF and ACF combinations for tensor accelerators.

[1]With sparse workloads, processing elements (PEs) that perform MAC operations of zero valued elements are considered underutilized.

TABLE I: MCF and ACF Characterization of SOTA accelerators.

| Design | MCF, ACF | Same? | Conv | E.g. |
|---|---|---|---|---|
| **Fix_Fix_None** | Fix, Fix | Yes | None | [4], [14], [15] |
| **Fix_Fix_HW** | Fix, Fix | No | HW | [16], [17], [18] |
| **Fix_Flex_HW** | Fix, Flex | No | HW | [19] |
| **Flex_Fix_SW** | Flex, Fix | No | SW | [20], [21] |
| **Flex_Fix_HW** | Flex, Fix | No | HW | [22], [23] |
| **Flex_Flex_None** | Flex, Flex | Yes | None | [5], [24] |
| **Flex_Flex_SW** | Flex, Flex | No | SW | CPU/GPU |
| **Flex_Flex_HW** | Flex, Flex | No | MINT | This work |

In fact, many accelerators are optimized for a certain workload and density region, and pick fixed choices for the MCF and ACF. In Table I we categorize how different formats and conversions happen in accelerators today[2]. Fix_Fix_None refers to accelerators that use a fixed and identical format for both MCF and ACF; therefore, no format conversion is needed. For e.g., the MCF and ACF of EIE [14] are both Dense(**A**)-CSC(**B**)-Dense(**O**). Fix_Fix_HW are accelerators that use different fixed MCF and ACF. For e.g., Eyeriss [17] stores and reads all fmaps in RLC. A hardware decoder transforms this RLC MCF into a dense ACF during computation. Flex_Fix_SW/HW is when the accelerator supports myriad choices for MCF but a fixed ACF. Software support relies on cuSPARSE [26] or Intel MKL [27]. Hardware support relies on a dedicated decompression unit. For e.g., NVDLA [22] has a dedicated ZVC to dense converter. Its MCFs are Dense(**A**)-ZVC(**B**)-Dense(**O**) and Dense(**A**)-Dense(**B**)-Dense(**O**), and its ACF is Dense(**A**)-Dense(**B**)-Dense(**O**). Flex_Flex_None refers to accelerators like Extensor [5], which supports multiple choices for MCFs and ACFs, but both must be the same, requiring no format converter.

In this work, we highlight two key challenges with using fixed format combinations (see Sec. III). First, the optimal format for compactness (i.e., MCF) might not be optimized for compute-efficiency (i.e., ACF), and vice versa. Second, the optimal format for compactness (or compute-efficiency) depends heavily on the tensor dimensions and sparsity ratio of the workload. As hardware accelerators in datacenters are expected to run a suite of scientific and DL applications, deployment of an accelerator optimal only for a certain sparsity region but inefficient for other regions is impractical and inefficient. We propose to extend tensor accelerators to support various compression formats (Flex_Flex_HW in Table I).

***The key contributions (Fig. 1b) of this paper are:***

- We analyze scientific and DL workloads across different MCFs and ACFs to understand how they perform at various sparsity regions (Sec. III). We make the case for accelerating kernels with a wide spectrum of sparsity ratios via three key extensions to current accelerators:
- First, we propose hardware extensions to accelerator PEs to support various ACFs to enhance utilization (Sec. IV).

[2]We would like to point out that while GPUs can technically support any MCF and any ACF, they lack hardware support for the specific choice of ACF, unlike custom accelerators. They are thus inefficient for sparse DL and scientific workloads. The scope of this work is on custom accelerators for tensor algebra and not GPUs, which are also adding sparsity support [25].

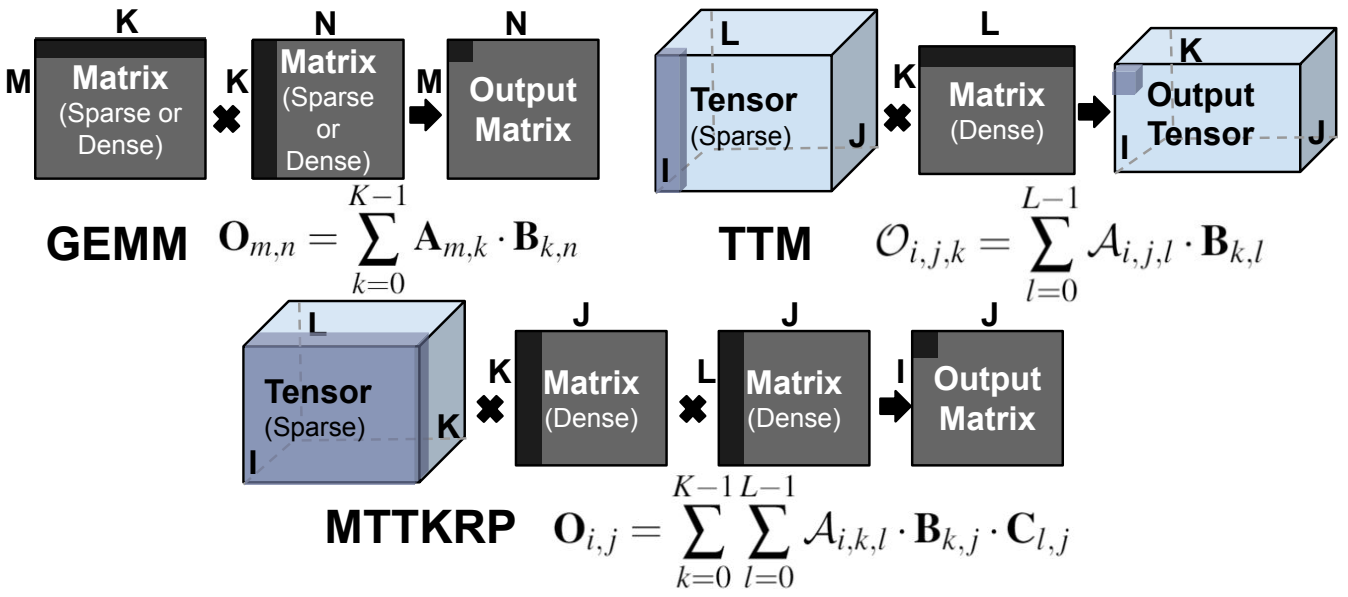

Fig. 2: Tensor kernels accelerated in accelerators [4], [5], [14], [24].

- Second, we present MINT[3], a hardware module that serves as a library of MCF to ACF converters (Sec. V). Rather than using separate converters, MINT reuses common building blocks across all conversions and also repurposes parts of the accelerator for format conversion.
- Third, we develop an analytical model called SAGE[4] to determine the optimized MCF and ACF combination given a workload, and configure MINT and the accelerator appropriately (Sec. VI).

## II. Background on Tensor Algebra

Tensors are multidimensional arrays used to represent the input and model datasets in DL and scientific workloads. Popular tensor kernels are shown in Fig. 2.

**GEMM.** General matrix-matrix multiplication (GEMM) along with its sparse matrix-sparse matrix multiplication equivalent (SpGEMM) and sparse matrix-dense matrix multiplication equivalent (SpMM) are often compute bottlenecks in both DL and scientific workloads. For DL, fully connected layers with batched inputs are rearranged into GEMMs. Convolutional layers are also commonly rearranged into GEMMs by using im2col or other efficient techniques [28], [29]. For scientific workloads and recommender systems, SpMM is often used for matrix factorization. SpMM and SpMV (sparse matrix-dense vector multiplication) are the key computational kernels in an iterative solver for sparse linear systems, while SpGEMM dominates the setup times of applications that use multigrid methods [30].

**TTM.** Sparse tensor times dense matrix multiplication (SpTTM) is a standard building block for all tensor computations. Tucker decomposition algorithm (often used for analyzing scientific datasets) intensively uses SpTTM, which is one of the main compute bottlenecks [31], [32].

**MTTKRP.** Matricized tensor times Khatri-Rao product (MTTKRP) is a core computation for canonical polyadic decomposition (CPD) algorithm and is the main compute bottleneck [32]. The three-dimensional tensor MTTKRP operation is shown in Fig. 2. Typically the tensor, $\mathcal{A}$, is sparse; while the matrices, **B** and **C**, are dense.

**Sparsity Formats.** Fig. 3 shows common lossless compression formats given a sparse matrix of size 4×4 and a sparse tensor of size 4×4×4. Popular formats in the scientific computing domain include COO, CSR, and CSC [8]. COO stores the nonzeros with their corresponding row and column locations. Rather than storing the row location like COO, CSR has *row_ptrs* that point to the corresponding *values* locations where each row starts. BSR is a blocked representation of CSR [33]. Given that the nonzeros follow a pattern, BSR reduces the metadata overhead and enables a more regular memory access pattern. Other formats that reduce metadata by clustering locations include HiCOO [10] and DIA [8]. CSF [11] constructs a tree to hold tensors. ZVC [9] stores nonzero elements along with a string of bits to represent each element (a bit value of 1 for a nonzero element and a bit value of 0 for a zero valued element). RLC [17] alternates between a nonzero element and the number of zeros between nonzeros.

[3]**M**icroarchitecture for **I**nterchangeable compressio**N** formats for **T**ensors
[4]**S**parsity form**A**t **G**eneration **E**ngine

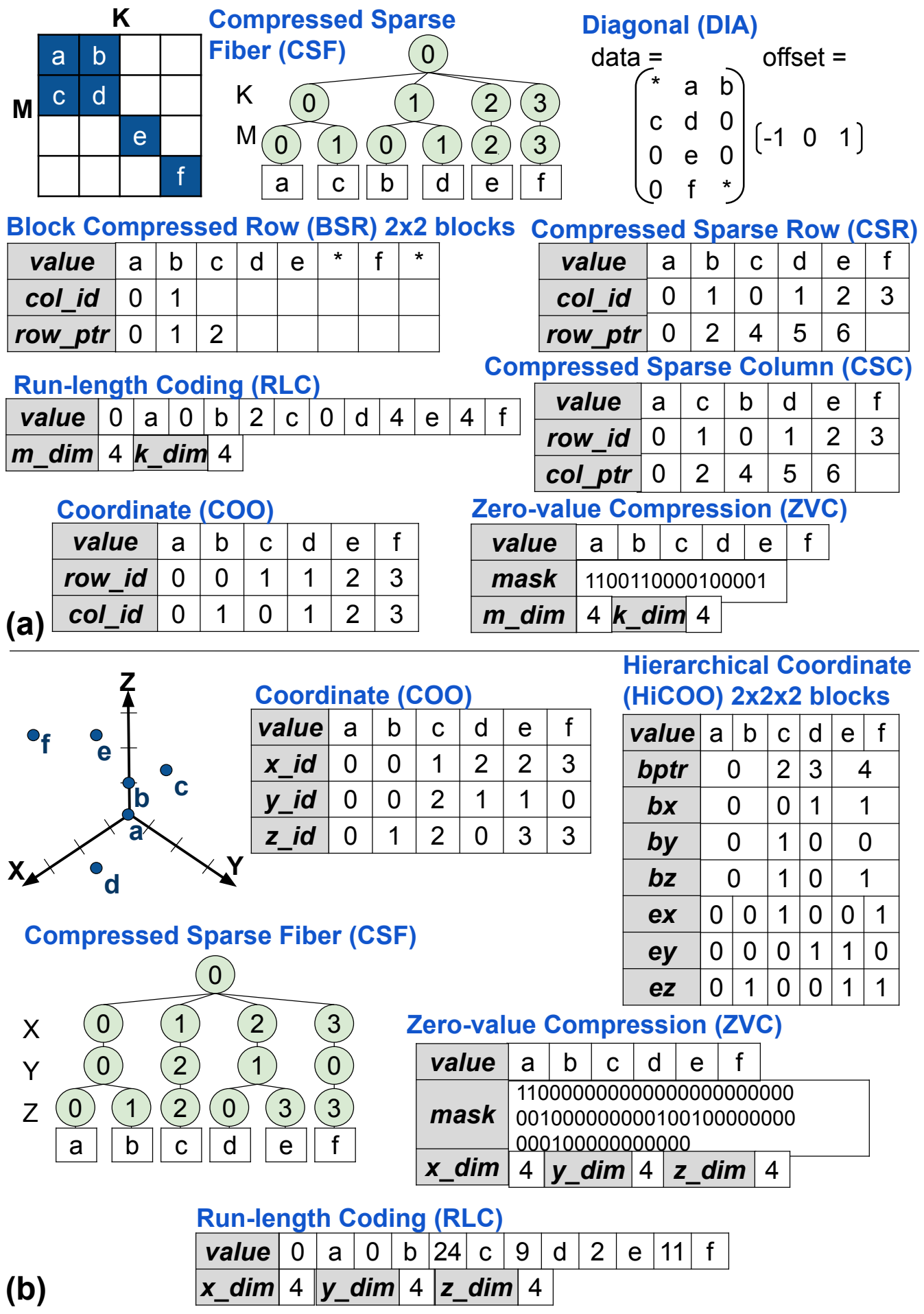

Fig. 3: Common compression formats for (a) matrix operations and (b) tensor operations. Not shown is Dense (uncompressed).

## III. Motivation: Compression Format Analysis

This section analyzes the performance of different compression formats and makes a case for supporting multiple unique MCFs and ACFs in accelerators. Recall from Sec. I that MCF is the format for memory storage and ACF is the format used for processing the algorithm. Note that many kernels require two or more tensor operands, and each operand can have its own MCF/ACF. An ideal MCF enables high *compactness* and an ideal ACF enables high *compute efficiency*.

### A. MCF Compactness Analysis

We assume a scratchpad in the accelerator for our compactness analysis. The number of metadata bits required is the log of the maximum possible value. Fig. 4a shows the relative energy cost of transferring a matrix of size 11k $\times$ 11k, with each compression format normalized to CSR. From the graphs, we observe that different compression formats are better at different sparsity regions. The regions to the left of the first red line in all Fig. 4a graphs indicate when CSR becomes more efficient than ZVC. For extreme sparsity, COO becomes the most compact format. The regions to the right of the second red line in all Fig. 4a graphs indicate when dense (uncompressed) becomes more efficient than CSR; while the middle region is well suited for RLC and ZVC. The figure also shows how quantization affects the effectiveness of each compression format. As the number of bits per data element goes down, the percentage of memory that goes to the compression format metadata goes up. Fig. 4b compares formats for extremely sparse matrices using a 16 bit datatype. Fig. 4bi, in particular, shows how increasing a dimension can affect the compactness of a format. The various metadata-to-data ratios make different MCFs outperform others.

***In summary, the compactness of a format depends on (1) tensor size, (2), sparsity region, and (3) datatype. There is no 'best' MCF.***

### B. ACF Performance Analysis

Different compression formats enable different ways of computing the desired output tensor. For e.g., Alg. 1 [34] shows a SpMM example with an input sparse matrix **A** in COO format, and the other matrix **B** in dense format. The algorithm iterates over the nonzero elements in the matrix **A** (*nnz*), and then multiplies with the corresponding elements in the matrix **B**. The generated output matrix **O** is in dense format; hereafter, this ACF is COO(**A**)-Dense(**B**)-Dense(**O**).

**Algorithm 1:** SpMM in COO format

**Input:** row_id, col_id and data: COO data structures for storing M$\times$K sparse matrix **A**, *nnz*: number of nonzero elements in **A**, **B**: K$\times$N dense matrix, **O**: M$\times$N dense output matrix

```
Initialize output matrix O to 0;
for i ← 0 to nnz − 1 do
    for j ← 0 to N − 1 do
        rid ← row_id[i]; cid ← col_id[i]; val ← data[i];
        O[rid][j] ← O[rid][j] + val ∗ B[cid][j];
```

To show how different ACFs result in varying performance, we benchmarked NVIDIA's cuBLAS and cuSPARSE libraries across all density regions. Fig. 5 shows the performance of four algorithms that use varying ACFs. Note that for this study, matrices **A** and **B** are set to the same density region. From Fig. 5a, we observe that Dense(**A**)-Dense(**B**)-Dense(**O**) performs better in density regions from 10% to 100%. CSR(**A**)-CSR(**B**)-CSR(**O**) performs better from $10^{-6}$% to 0.1%. Fig. 5b and Fig. 5c show the SM and memory utilization of Fig. 5a.

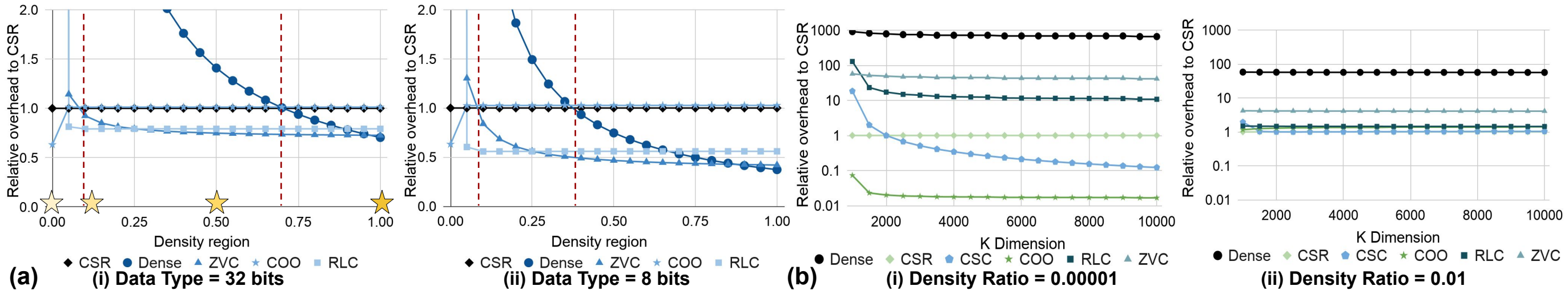

Fig. 4: (a) Relative energy cost of transferring an 11k $\times$ 11k compressed matrix from DRAM with varying compression formats, density regions, and datatypes. The stars highlight density regions of $10^{-6}$%, 10%, 50% and 100%. (b) Relative energy cost of transferring extremely sparse matrices from DRAM with varying compression formats and K dimensions. M dimension is set to 1k.

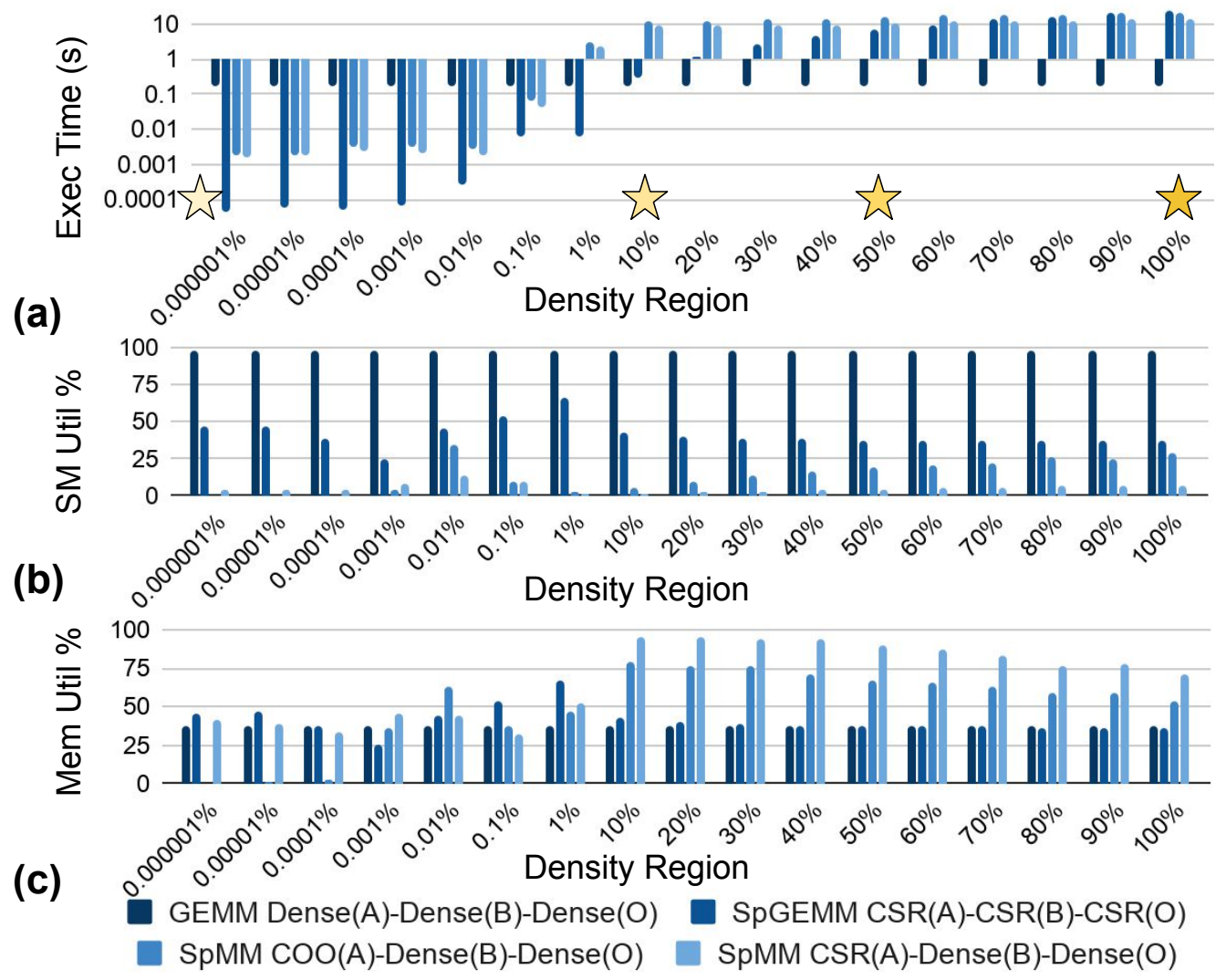

Fig. 5: Performance of MM algorithms (different ACFs) from cuBLAS and cuSPARSE on Titan GPU across density regions. (a) Exec time, (b) SM util. and (c) Mem util. with M, N, and K of size 11k. The stars are density regions of $10^{-6}$%, 10%, 50% and 100%.

GEMM is compute bound, but note that SM utilization includes zero-valued operations. SpGEMM is often latency bound, while the other two SpMM algorithms are often memory bound. How much resources are spent processing the metadata vs computing the data enable different ACFs to outperform others at different sparsity regions. Though not shown, we observe that changing the tensor size affects the relative performance of each ACF.

***In summary, the compute efficiency depends on (1) tensor size and (2) sparsity region. There is no 'best' ACF.***

Fig. 4a-i and Fig. 5a show stars for four specific density regions: $10^{-6}$%, 10%, 50% and 100%. Both are measured using matrices of 11k by 11k with float32 datatype. At $10^{-6}$%, COO is the best MCF, while CSR(**A**)-CSR(**B**)-CSR(**O**) is the best ACF. RLC, ZVC and dense (uncompressed) are the best MCF for 10%, 50% and 100% density region respectively; while Dense(**A**)-Dense(**B**)-Dense(**O**) is the best ACF for all three.

***In summary, the most compact MCF and the most compute efficient ACF need not be the same.***

### C. Other Use Cases For Format Conversions

Conversion between different compression formats not only result in better *compactness* and *compute efficiency* as shown above, but can also be necessary for algorithmic reasons. For instance, during backpropagation in DL training, converting CSR to CSC (or vice versa) is necessary since the weight matrix gets transposed before running GEMM. Similarly, accelerators that output in dense format (e.g., TPU) may require compression before storing back to memory.

## IV. Accelerator Microarchitecture Extensions

As discussed in Sec. III, different formats are ideal for different datatypes, tensor sizes, and density regions. This motivates the need for supporting multiple sparsity formats within the accelerator. In this section, we present the extensions required to support various ACFs over an example accelerator via the walkthrough example in Fig. 6. In Sec. V and Sec. VI we present a format converter and format predictor respectively.

### A. Accelerator Architecture

We assume an accelerator template consisting of an array of PEs connected to a global shared scratchpad buffer via specialized network-on-chip (NoC), similar to prior works [14], [15], [18], [19]. Each PE has a vector MAC unit, weight buffer, output register, and state registers. For our walkthrough example, we assume a weight stationary (WS) dataflow, i.e., elements of matrix B stay stationary at the PEs, while matrix A is streamed. In addition, PEs in sparse accelerators [14], [15], [18] house an indexing unit to match the non-zero input-weight pairs that need to be multiplied.

**Microarchitecture Extensions.** To enable sparse accelerators to support multiple ACFs, we propose two microarchitecture extensions. The first is to enable flexible resource allocation of scratchpad buffer space within each PE for metadata (color red in Fig. 7) and stationary data (color green in Fig. 7); i.e. a buffer entry can be treated as either data or metadata depending on a flag. This is essential because different ACFs have different ratios of metadata. The PEs also need comparators for metadata index matching like any other sparse accelerator [14], [15]. The second extension is to have flags next to the data being delivered over the bus/NoC by the global buffer to identify whether it is the operand data or metadata for the format.

To find matching indices, metadata from the buffer is compared with the metadata from the input bus/NoC. The outputs of the comparators go to a one-hot-to-binary encoder to find the location of the valid data within the data region. Then, the matching data from the input bus and buffer is sent to a vector unit. Fig. 7 shows that the extension increases the size of a PE with 128B buffer by ∼10%. We use a PE with vector size of eight 32-bit compute units.

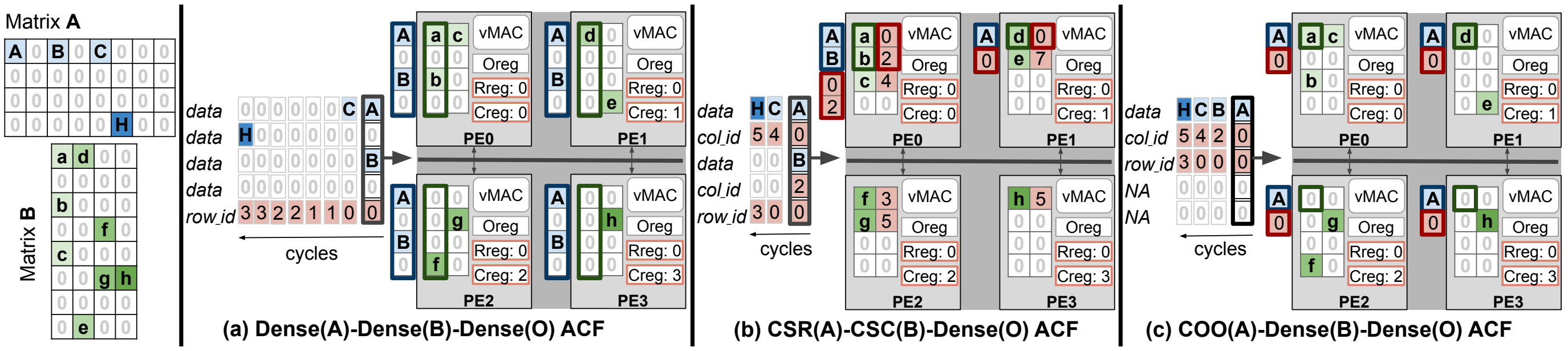

Fig. 6: Examples of different ACFs mapped onto a sparse weight-stationary accelerator. Each column of Matrix **B** is loaded stationary to each PE. Depending on Matrix **B**'s ACF, many zero-valued elements (e.g Dense ACF) or metadata (colored red, e.g CSC ACF) consume buffer space. Next, Matrix **A** gets sent through a broadcast bus to all PEs, and matches corresponding elements for multiplication. Similarly, depending on Matrix **A**'s ACF, the bus may contain many zero-valued elements (e.g Dense ACF), or metadata (colored red, e.g. CSR and COO ACF). The key takeaway is that ACFs affect both buffer utilization and data streaming latency.

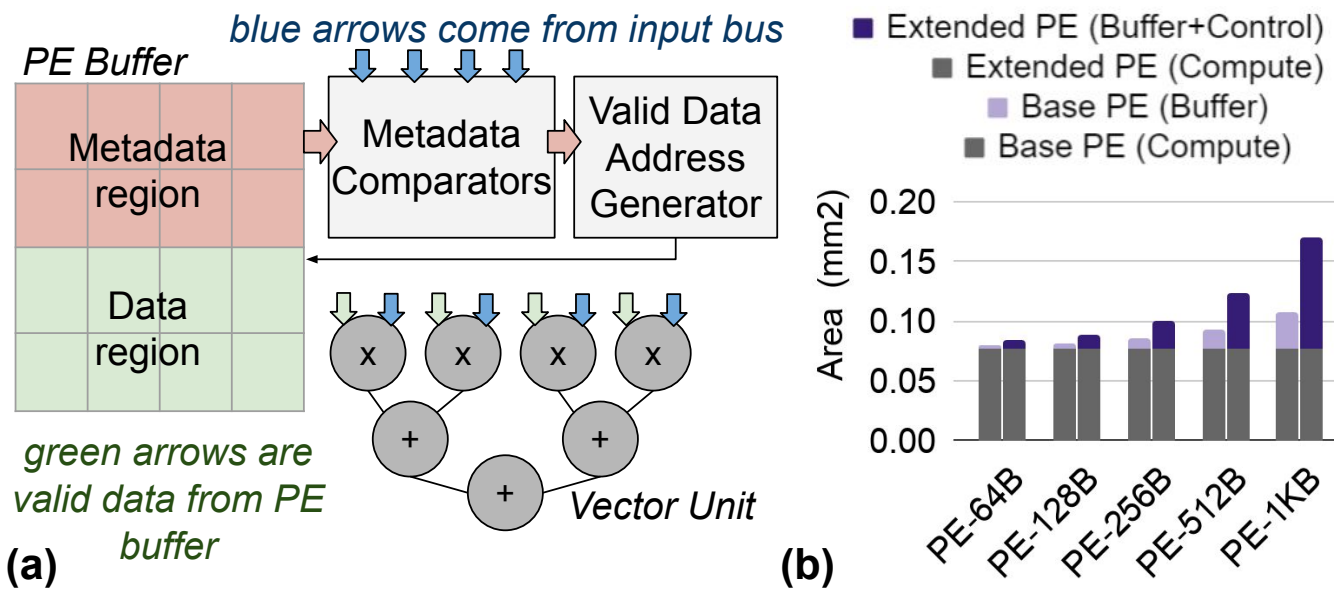

Fig. 7: (a) Extended PE microarchitecture. (b) Overhead of extended PE compared to base PE.

### *B. Walkthrough Example.*

In Fig. 6, we assume 4 PEs, a distribution bandwidth of five elements per cycle, and a weight buffer size of eight elements per PE. Each PE has a output row register (Rreg), output column register (Creg), and output value register (Oreg). We assume that each metadata and data element consume the same amount of resources (bandwidth and buffer space). Note that differently sized partitions are needed if the metadata-to-data ratio is not one-to-one.

Fig. 6a shows Dense(**A**)-Dense(**B**)-Dense(**C**) ACF. Each column of matrix **B** (weights) is stored stationary within each PE and Creg is set to the column number. Since the matrix **B** is in dense format, the zeros must be stored to maintain correct buffer indexing. Four *data* elements and one *row_id* from matrix **A** are broadcasted to all PEs per cycle. The *row_id* is generated by a controller and is used to set the Rreg. Rreg and Creg are used track the output index. The streaming (blue) data gets matched to the corresponding stationary (green) data. Both data are then forwarded to the vector MAC units for computation. The partial product is stored in Oreg and continues to accumulate locally until either Creg or Rreg changes. Once that happens, Oreg gets sent to a global output buffer. The address where Oreg goes to is determined by Creg and Rreg.

Fig. 6b shows CSR(**A**)-CSC(**B**)-Dense(**C**) ACF. Since the weights are stored in CSC, the buffers include the nonzero elements and their corresponding row indices. Therefore, half of the buffer is allocated to metadata entries. The input bandwidth consists of two *data* elements, two *col_ids*, and one common *row_id*. However, if the *row_id* is not common among both *data*, it must be broken up as shown by 'C' and 'H'. In the first cycle at PE0, streaming elements 'A' and 'B' get matched with stationary elements 'a' and 'b'. This occurs because the streaming *col_ids* of '0' and '2' match with the stationary row indices of '0' and '2'.

Fig. 6c shows COO(**A**)-Dense(**B**)-Dense(**C**) ACF (refer to Alg. 1). For each nonzero element in the matrix **A**, *data*, *col_id*, and *row_id* are broadcasted to each PE. Using the *col_id*, the PEs access the correct weight buffer entry to get the corresponding data for multiplication. For e.g., at PE0, because the *col_id* is '0', element 'A' matches with the element at buffer index '0'. Notice that only one data entry (color blue) can be sent per cycle. This is because there is not enough bandwidth to send two at a time, as (2 × (1 data + 2 metadata)) > (bandwidth size of 5). Overall Fig. 6a,b,c require 8, 3, and 4 cycles to send matrix **A** respectively; showing how different ACFs affect the overall performance.

## V. MINT: Compression Format Converter

To support multiple MCFs and ACFs in a single accelerator (Sec. III), we propose MINT, a general purpose format converter next to the accelerator, instead of offloading conversions to the host CPU /GPU as accelerators do today. Given $m$ MCF's and $a$ ACF's supported by the architecture, MINT can provide $m \times a$ conversions for any combination of MCF and ACF deemed optimal for the current workload by SAGE (Sec. VI). MINT's efficiency comes from (i) merging building blocks to one general-purpose converter rather than having $m \times a$ separate converters, and (ii) reusing and repurposing compute modules from the accelerator's datapath for the format conversion.

### *A. MINT Designs and Building Blocks*

Fig. 8a shows the different MINT designs and building blocks. The building blocks are composed of prefix sum, memory controller, parallel divide, parallel mod, comparators, and other format specific modules. There are three different MINT implementations: MINT baseline (MINT_b), MINT merge (MINT_m), and MINT merge + reuse (MINT_mr).

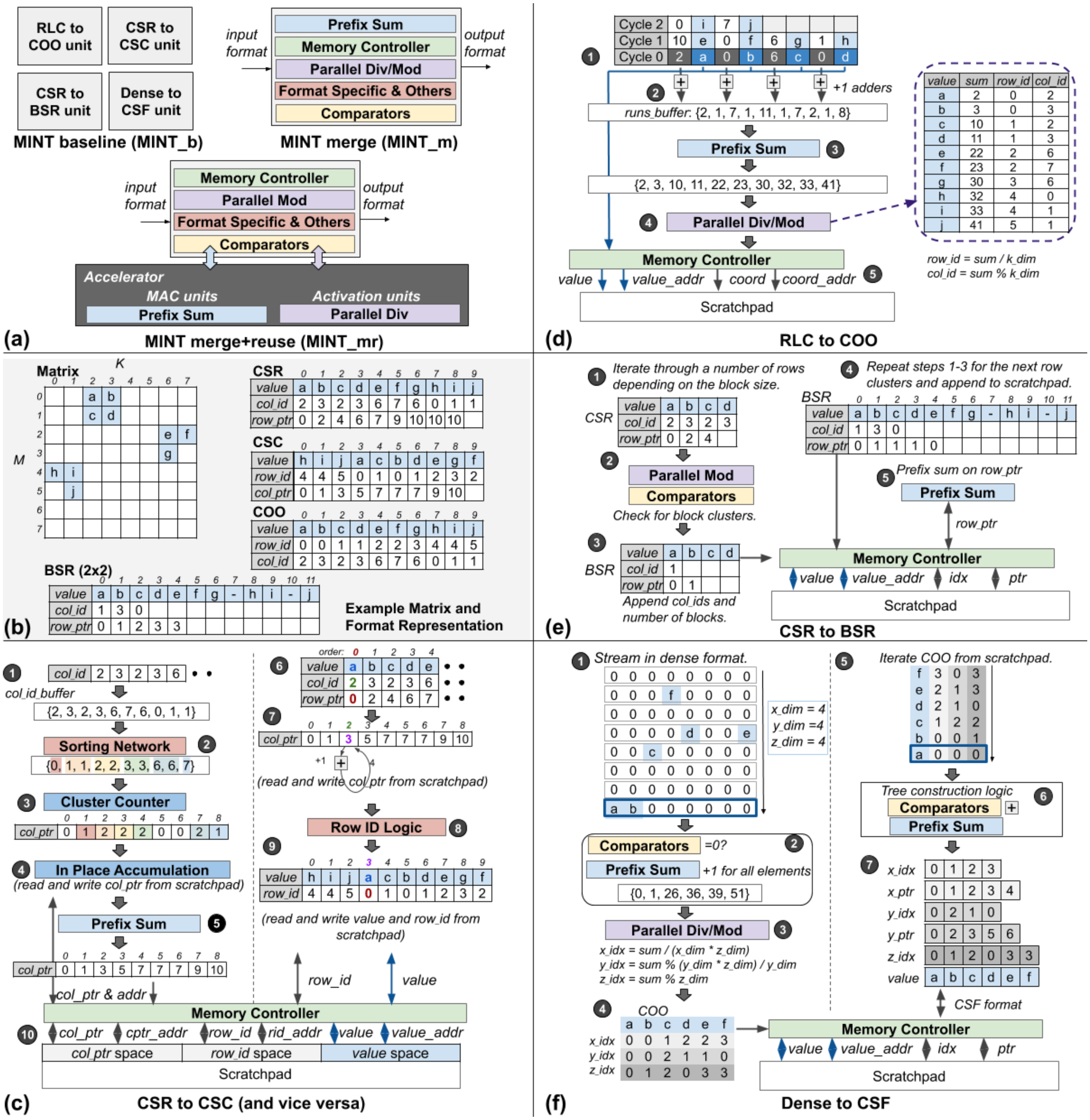

Fig. 8: (a) High level MINT designs with building blocks for format conversion. (b) Matrix used in example c-e. (c) CSR to CSC conversion. (d) RLC to COO conversion. (e) CSR to BSR format. (f) Dense to CSF conversion using the tensor from Fig. 3b.

MINT_b composes of separate format converters. MINT_m generalizes overlapping building blocks and merges them together. MINT_mr expands on MINT_m by reusing the MAC (for prefix sum) and activation units (for parallel divides) from the accelerator, as we discuss next.

**Adders for Prefix Sum.** Prefix sums are often used during format conversions. Specific examples are discussed in Sec. V-B. All accelerators that target tensor kernels already have adders inside their PEs [4], [5], [16], [18], [24], which we repurpose to compute prefix-sums.

There are three main implementations of prefix sums: (a) serial chain, (b) work efficient, and (c) highly parallel designs, as shown in Fig. 9. The throughput-optimized serial chain implementation can reuse a store-and-forward spatial reduction network by adding diagonal links across adders as shown in Fig. 9a. Registers are used to maintain timing. To solve blocking issues and to maintain a throughput of N prefix sum outputs in a cycle, a row of adders at the end is needed to add the offset, which is the maximum value from the previous cycle. Work efficient prefix sums can reuse the adders in an adder tree reduction network as shown in Fig. 9b. Highly parallel prefix sums achieve a latency of logN cycles but require more active adders and forwarding links as shown in Fig. 9c.

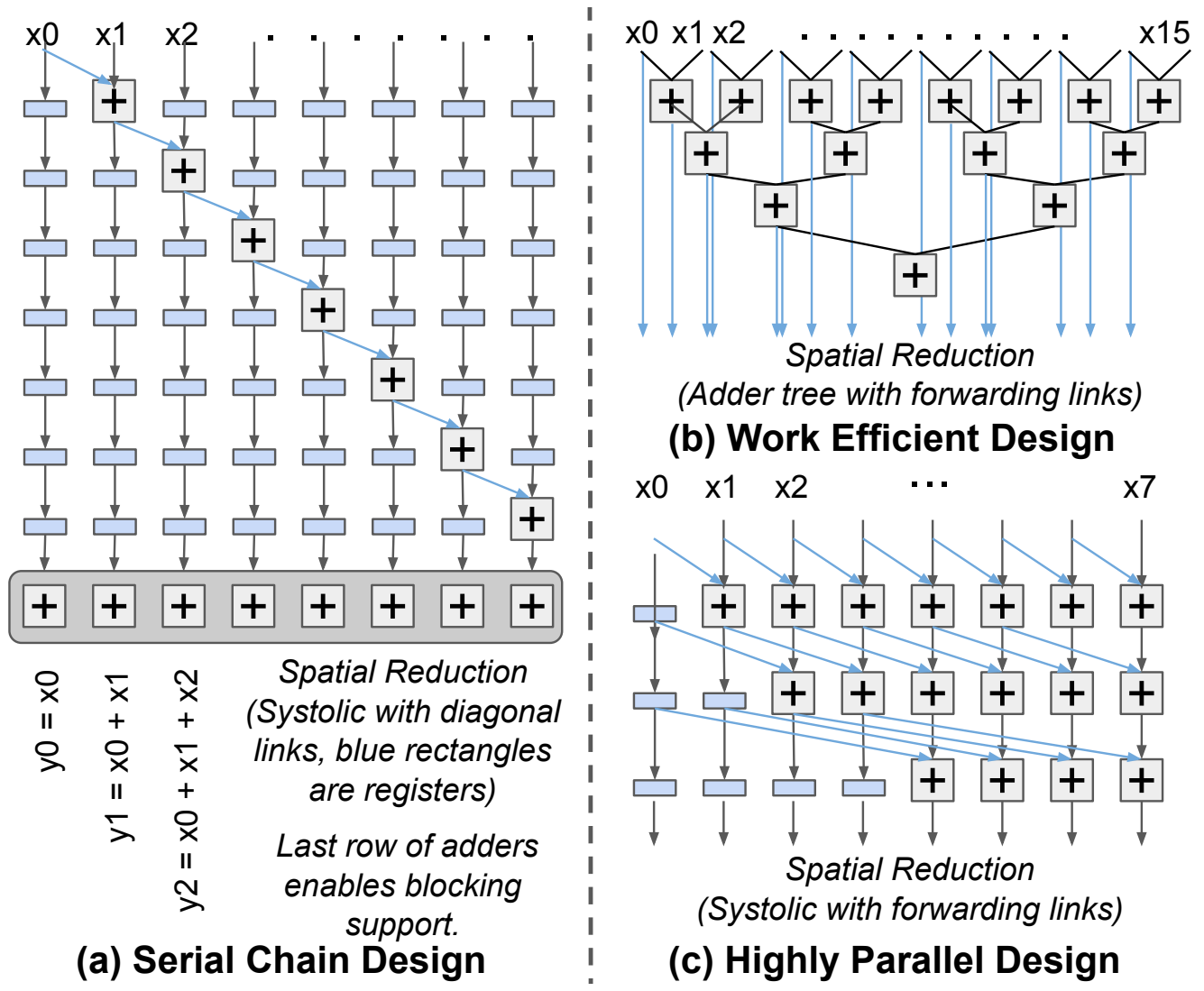

Fig. 9: Common prefix sum (scan) implementations and how they can be implemented on top of accelerator hardware structures.

There are two requirements to enable prefix sums in existing accelerators. The first requirement is that the accelerator must support int32 adds (often used for metadata calculations). The second requirement is that the accelerator need wire and mux modifications. We observe that the overhead for enabling these modifications is minimal in the evaluation section.

**Activation Units for Position Calculations.** DL accelerators often have activation units attached [4], [22]. They contain varying types of compute units such as dividers, which support many DL operations including softmax and batch normalization. They can also be used to calculate metadata positions, enabling potential compute reuse (refer to Sec. V-B).

### B. *Example Conversions*

To discuss how format conversions are implemented in MINT, we select four representative conversions that are typically used in tensor accelerators. CSR to CSC is useful for transposing weights during DL backpropagation. RLC to COO is useful because RLC is a common MCF, while COO enables fast translation to other formats. CSR to BSR is common for accelerators that benefit from operating on structured data. Dense to CSF is useful as many accelerators output in dense format; and compression may be beneficial before storing back to memory. Fig. 8 shows how the four conversions can be broken down into building blocks. Other format conversions can be generalized with the same building blocks; For e.g., ZVC-to-Dense and Dense-to-ZVC [9].

*1) CSR to CSC:* Fig. 8b shows the target matrix and Fig. 8c presents the CSR to CSC conversion. ❶ reads a chunk of the *col_ids* and stores them a buffer. ❷ sorts the chunk and ❸ counts the number of specific values within the chunk. It is also possible to stream the *col_ids* and atomically increment the cluster counter. Note that steps 2 and 3 are not needed if the size of the chunk is one. ❹ allocates space in the scratchpad to read and write the *col_ptr*. This continues until all of the input *col_ids* are read and accumulated to the corresponding location in the *col_ptr*. ❺ The entire *col_ptr* goes through a prefix sum module. ❻ iterates across the CSR fields. The first entry uses the *col_ids* (2) to index the *col_ptr*, which contains the *value* (3). This is used to index the *value* and *row_id* of the CSC format. ❼ After getting the *value* (3), the *col_ptr* increments by one for the next reference. For example, the *value* 'a' gets written to location 3. Three gets updated to four, so that *value* 'c' gets written to location 4 in the future. This is to prevent overwriting data. ❽ contains logic to determine the output *row_id* by counting the number of iterated values versus the *row_ptr*, which shows how many elements are in every row. ❾ shows the final updated values and *row_ids* of the CSC format, while ❿ stores the outputs into the correct memory space. The *col_ptr* is fixed slightly during this step to point the original values.

*2) RLC to COO:* Fig. 8b shows the target matrix. Fig. 8d presents the RLC to COO conversion. ❶ shows the matrix compressed in RLC format being streamed in. ❷ adds one to all elements beside the first element of cycle 0. This is to offset the position of the level (nonzero value). ❸ Depending on the buffer size, all elements go to a prefix sum module. ❹ After generating the prefix sums, the outputs go into a parallel divide and mod units. The *row_ids* is generated by dividing the sum with the K dimension of the matrix, and the *col_ids* is generated by moding the sum with the K dimension. ❺ The memory controller stores the level fields (nonzero elements) from step 1 and the coordinates from step 4 to the allocated memory spaces.

*3) CSR to BSR:* Fig. 8b shows the target matrix. Fig. 8e shows the conversion of CSR to $2\times2$ BSR. ❶ iterates through the first two rows of the matrix (which we refer to as a row block). ❷ Then, mods and comparators are used to find the block position and to determine whether the block has already been initialized or not. ❸ appends the number of unique blocks to the *row_ptr*, while the column position of the block goes into the *col_ids*. Register flags are used to keep track of initialized blocks so elements that fit within the block region are written to the correct place in memory. Note that zeros are inserted into the values if the blocks are not complete. (CSR does not contain any zero values, while BSR may contain zero values based on the completeness of the block structure.) ❹ repeats steps 1 - 3 for the remaining row blocks. ❺ After iterating through the whole matrix, the *row_ptrs* are read from the scratchpad via the memory controller and fed into a prefix sum module.

*4) Dense to CSF:* Fig. 8f shows the conversion from Dense to CSF and refers to the tensor from Fig. 3b. ❶ shows the dense format equivalent in z $\rightarrow$ y $\rightarrow$ x order from bottom-left to top-right. Note that different orders are valid for dense format. The exact order must be specified to the decoder beforehand.

TABLE II: MCF and ACFs for SOTA Accelerators Evaluated. The naming scheme is MCF_ACF_Converter.

| Type | MCF (A-B) | ACF (A-B) | E.g. |
|---|---|---|---|
| **Fix_Fix_None** | Dense-Dense | Dense-Dense | TPUv1 [4] |
| **Fix_Fix_None2** | CSR-Dense<br>Dense-CSC | CSR-Dense<br>Dense-CSC | EIE [14] |
| **Fix_Flex_HW** | ZVC-ZVC | CSR-Dense<br>Dense-CSC<br>Dense-dense | SIGMA [19] |
| **Flex_Flex_None** | (CSR/Dense)-(Dense/CSC) | (CSR/Dense)-(Dense/CSC) | ExTensor [5] |
| **Flex_Fix_HW** | (ZVC/Dense)-(ZVC/Dense) | Dense-Dense | NVDLA [22] |
| **Flex_Flex_SW** | CSR-Dense | CSC-CSR | Intel MKL cuSPARSE |
| **Flex_Flex_HW** | From SAGE | From SAGE | This work |

❷ checks if the elements in dense format are zero or not, and the prefix sum module adds one to each element of the streaming dense format. ❸ contains parallel dividers and mod units to get the COO coordinates of the nonzero values. The equations are shown in the figure. *x_idx*, *y_idx*, *z_idx* represent the COO coordinates. The sum represents the prefix sum value of that particular nonzero element. While *x_dim*, *y_dim* and *z_dim* represent the corresponding dimension of the tensor. ❹ shows the generated COO format equivalent of dense format, and ❺ iterates COO to begin generating CSF. ❻ includes the tree construction logic which consists of comparators, adders, and a prefix sum unit. ❼ contains the final CSF structure that is located in the scratchpad.

*In summary, the compute components for format conversions can be broken down into building blocks; enabling hardware reuse for many other conversion combinations. For performance, MINT is pipelined to start conversion while streaming in data from memory.*

## VI. SAGE: Compression Format Predictor

SAGE predicts which MCF and ACF combination results in the lowest energy-delay product (EDP). The inputs to SAGE are workload size, datatype, density region, MINT format conversion cost, and accelerator hardware parameters (Fig. 1b). The outputs are the ideal MCF and ACF combinations. Note that there might be scenarios when the MCF is already predetermined by the programmer. In that case, the SAGE will find the best accelerator configuration (ACF) and conversion type. SAGE contains a cost model and performance model.

**Cost Modeling.** The cost model first predicts the DRAM energy consumption and transfer cycles cost. This is directly proportional to the compression size of the MCF. Second, to model the conversion cost, we evaluate the building blocks necessary for each conversion scenario along with their relative execution cycles and power consumption.

**Performance Modeling.** The performance model assumes a WS accelerator, described in Sec. IV, and a flexible NoC to deliver non-zeros from the streaming tensor [5], [19]. It estimates the buffers used by data and meta-data (assuming full flexibility within the scratchpads for this partition), and calculates the number of compute cycles for all target ACFs (similar to Fig. 6). For the target matrices and tensors, we assume a uniform random distribution of the dense values. This has minimal effect on the performance of unstructured format conversions, because the number of calculations is directly proportional to the tensor dimensions and the number of nonzeros. However, assuming random sparsity does not apply to structured formats. Enhancing the performance model for structured formats (e.g. DIA, HiCOO, BSR and ELLPACK [8]) is part of our future work.

## VII. Evaluation

### A. Methodology

**Accelerators.** We contrast our proposed accelerator with flexible MCF-ACF support against a suite of SOTA sparse accelerators shown in Table II. All accelerators are given 16384 total MAC units (similar to Google TPU [4]), 512B of buffer storage per PE, 512-bit input bus per cycle, and 32-bit datatype.
**Workloads.** For our evaluations, we use matrices and tensors from the following datasets: Suitesparse [1], Deepbench [35], FROSTT, [3], and BrainQ [36]. Deepbench represents DL workloads and the others represent scientific workloads. The dimension, number of nonzeros, and density ratio characteristics are shown in Table III.
**Compression Formats.** For MCF, we consider six format choices for each operand: Dense, RLC, ZVC, COO, CSR, and CSC. For ACF, we consider four format choices for each operand: Dense, COO, CSR, and CSC. Table III shows the ideal MCF and ACF combinations for each workload, determined by SAGE(Sec. VI). Combinations shaded blue are ideal for SpGEMM, grey for SpMM, tan for SpTTM, and yellow for MTTKRP. The factorizing matrices that are multiplied with the tensors are generalized to have dimensions of K by (M/2).

### B. MINT Performance and Cost

**MINT vs Software.** We contrast the performance of MINT (i.e., Flex_Flex_HW) versus the time it would take to do format conversion using software on a CPU and GPU (i.e., Flex_Flex_SW). For CPU evaluations, we use the Intel's MKL library to evaluate Dense to CSR and CSR to CSC. We use an Intel Core i9-9820X CPU operating at 3.3GHz. It has 10 cores, 85GB/s memory bandwidth and a TDP of 165W. For GPU evaluations, we use cuSPARSE on an NVIDIA Titan RTX GPU operating at 1.77GHz. It has 4,608 CUDA cores, 672GB/s memory bandwidth and a TDP of 280W. Fig. 10 compares the wall time among Intel's MKL, NVIDIA's cuSPARSE, and MINT for CSR to CSC and Dense to CSR. Note that few conversions are compared because of limited library support. MINT shows faster average conversion time than both CPUs and GPUs. This is because MINT is able to efficiently overlap conversion with data streaming from memory. Since MINT is a hardware module design placed next to the accelerator, it also observes roughly three orders of magnitude energy improvement. Besides performing better compute-wise, MINT offers significant speedup and energy savings by reducing the host to device (H2D) and device to host (D2H) transfer times. As shown in Fig. 11, transferring data can consume up to 75% of the total time, and has a geomean of roughly 50%. Thus, it is critical to have hardware support for format conversion.

TABLE III: Tensors for our evaluations (left side). MCFt is the format of the input tensor, and MCFf is the format of the input factorization matrix(es). Blue shaded formats are ideal MCFs and ACFs for SpGEMM. Grey shade for SpMM. Tan shade for SpTTM. Yellow shade for MTTKRP. The format combinations are generated by SAGE.

| Workload | Dimensions | #Nonzeros | Density% | MCFt | MCFf | ACFt | ACFf | MCFt | MCFf | ACFt | ACFf |
|---|---|---|---|---|---|---|---|---|---|---|---|
| *journal* [1] | 124 × 124 | 12K | 78.5 | ZVC | ZVC | Dense | Dense | ZVC | Dense | Dense | Dense |
| *bibd* [1] | 171 × 92K | 3.3M | 20.9 | RLC | CSC | Dense | CSC | RLC | Dense | Dense | Dense |
| *dendrimer* [1] | 730 × 730 | 63K | 11.8 | RLC | CSC | Dense | CSC | RLC | Dense | Dense | Dense |
| *speech1* [35] | 11K × 3.6K | 3.9M | 10.0 | RLC | CSC | Dense | CSC | RLC | Dense | Dense | Dense |
| *speech2* [35] | 7.7K × 2.6K | 1M | 5.0 | RLC | CSC | Dense | CSC | RLC | Dense | Dense | Dense |
| *nd3k* [1] | 9K × 9K | 3.3M | 4.1 | RLC | CSC | Dense | CSC | RLC | Dense | Dense | Dense |
| *cavity14* [1] | 2.6K × 2.6K | 76K | 1.1 | CSR | CSC | Dense | CSC | CSR | Dense | CSR | Dense |
| *model3* [1] | 1.6K × 4.6K | 24K | $3.2x10^{-1}$ | CSR | CSC | CSR | CSC | CSR | Dense | CSR | Dense |
| *cat_ears* [1] | 5.2K × 13.2K | 40K | $5.7x10^{-2}$ | CSR | CSC | CSR | CSC | CSR | Dense | CSR | Dense |
| *m3plates* [1] | 11K × 11K | 6.6K | $5.4x10^{-3}$ | COO | COO | CSR | CSC | COO | Dense | CSR | Dense |
| *BrainQ* [36] | 60 × 70K × 9 | 11M | 29.1 | ZVC | Dense | Dense | Dense | ZVC | Dense | Dense | Dense |
| *Crime* [3] | 6.2K × 24 × 2.5K | 5.2M | 1.5 | CSF | Dense | CSF | Dense | CSF | Dense | CSF | Dense |
| *Uber* [3] | 4.4K × 1.1K × 1.7K | 3.3M | $3.9x10^{-2}$ | COO | Dense | CSF | Dense | COO | Dense | CSF | Dense |

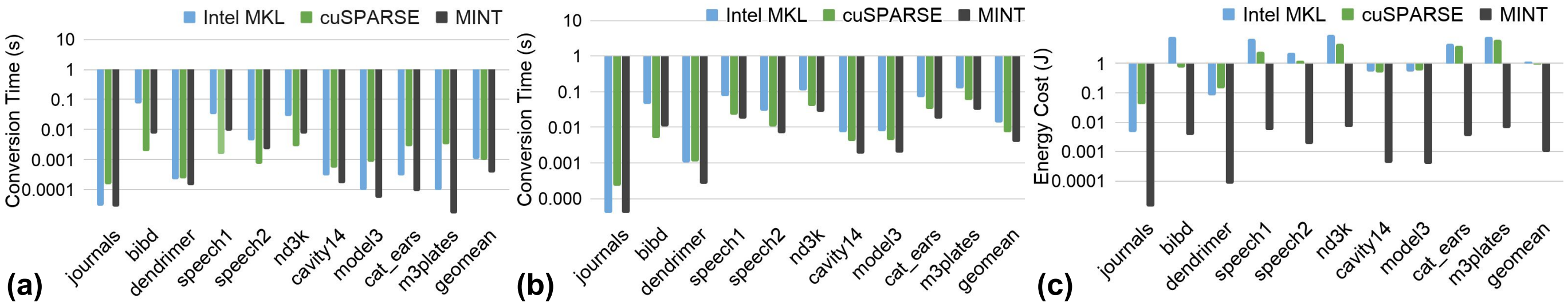

Fig. 10: (a) CSR to CSC conversion execution time. (b) Dense to CSR conversion execution time. (c) Energy consumption analysis.

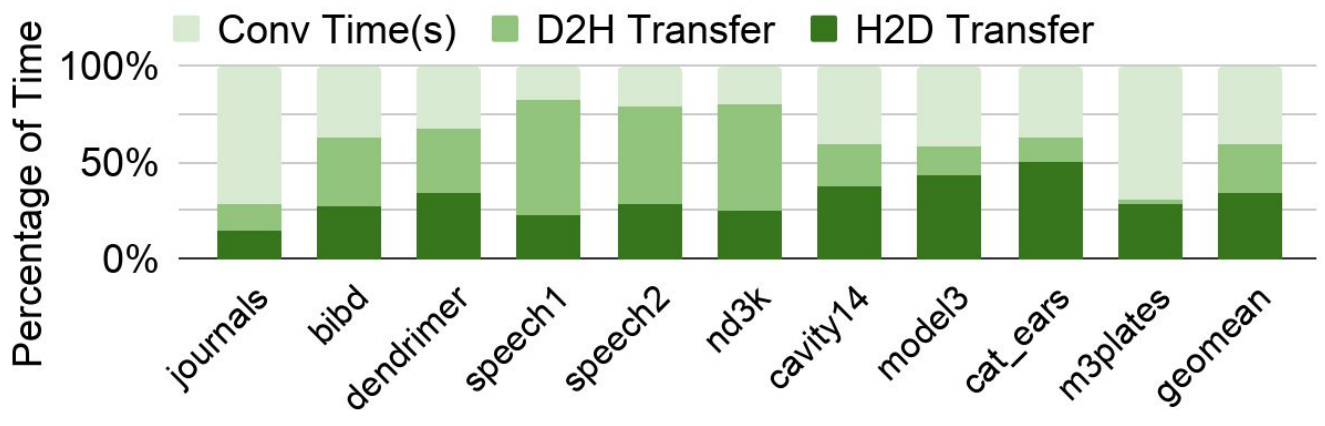

Fig. 11: Average GPU transfer to compute ratio, highlighting additional costs of offloading conversions rather than doing it locally.

**MINT Overhead.** To get area and power estimates, we implemented the building blocks of MINT in RTL and performed synthesis and place-and-route using a 28nm PDK at a clock rate of 1GHz. For our MINT implementation, we limit the number of parallel mod and divider units to eight due to how hardware expensive the modules are. Other components include a pipelined sorting network (input size equal to the number of unique metadata coming in per cycle), cluster counter, prefix sum unit, numerous comparators, eight multipliers, and a memory controller with address generators, FIFOs, and crossbar. Using the conversions in Fig. 8, MINT_b, MINT_m, and MINT_mr consumes 0.95 mm2, 0.41 mm2, and 0.23 mm2 respectively. MINT_m has a ∼57% area reduction over MINT_b from building block reuse, while MINT_mr has an additional ∼45% area reduction over MINT_m from reusing compute units found inside the host accelerator.

MINT_mr requires slight modifications to existing int32-supported accelerators. We enable highly parallel prefix sum of 32 inputs by modifying adders with forwarding links and muxes. The size of 32 is selected to satisfy MINT throughput requirement. We observe roughly a 20% increase in area and 27% increase in power. As shown in Fig. 9, other reduction networks in accelerators can be augmented to enable various versions prefix sum. Reusing the dividers in the activation units require a mux, controller, and dedicated data paths.

To reduce MINT_mr's area and power overhead, a serial chain prefix sum design (Fig. 9a) can be used instead of a highly parallel prefix sum design (Fig. 9c). Serial chain has a longer tail latency; but has simpler wiring, fewer muxes, and fewer active adders than highly parallel design. It is also possible to make the design smaller by reducing the number of parallel inputs in exchange for less throughput. To overlay a 16 × 16 int32 PE array with serial chain prefix sum, we observe a 2% increase in area and 3% increase in power. Divider and mod units must be pipelined to meet timing. Together, they consume 74% and 65% of MINT_m's area and power respectively. In comparison to a 16384 PE accelerator with (int16/int32 & bfp16/fp32) support, MINT_m consumes 0.5% of its area and 0.4% of its power overhead. Note that power gating is possible when no conversion is operating.

## C. Performance Breakdown

Fig. 12 shows the breakdown of SpGEMM on *journals*, *speech2*, and *m3plates* from Table III with different accelerator from Table II. With SAGE's ability to find the best MCF and ACF configurations, it is possible to achieve lower EDP than the other accelerators. Fig. 12a shows that Fix_Fix_None2 takes the most cycles and energy. This is because *journals* is relatively dense, so an ACF of Dense(**A**)-Dense(**B**) is better than Dense(**A**)-CSR(**B**). Flex_Fix_HW and our work is similar because both choose MCFs of ZVC(**A**)-ZVC(**B**) and ACFs of Dense(**A**)-Dense(**B**). Fig. 12b shows that Dense(**A**)-CSC(**B**) is the best ACF, as Fix_Fix_None2, Flex_Flex_None and our work have similar compute times. Since our design has MINT, it is able to use RLC as the MCF, reducing memory transfer time. Fig. 12c shows that CSR(**A**)-CSC(**B**) is the best ACF, as Flex_Flex_None and our work show much lower compute and energy cost. Since *m3plates* is extremely sparse, any ACF with dense format will lead to poor compute efficiency.

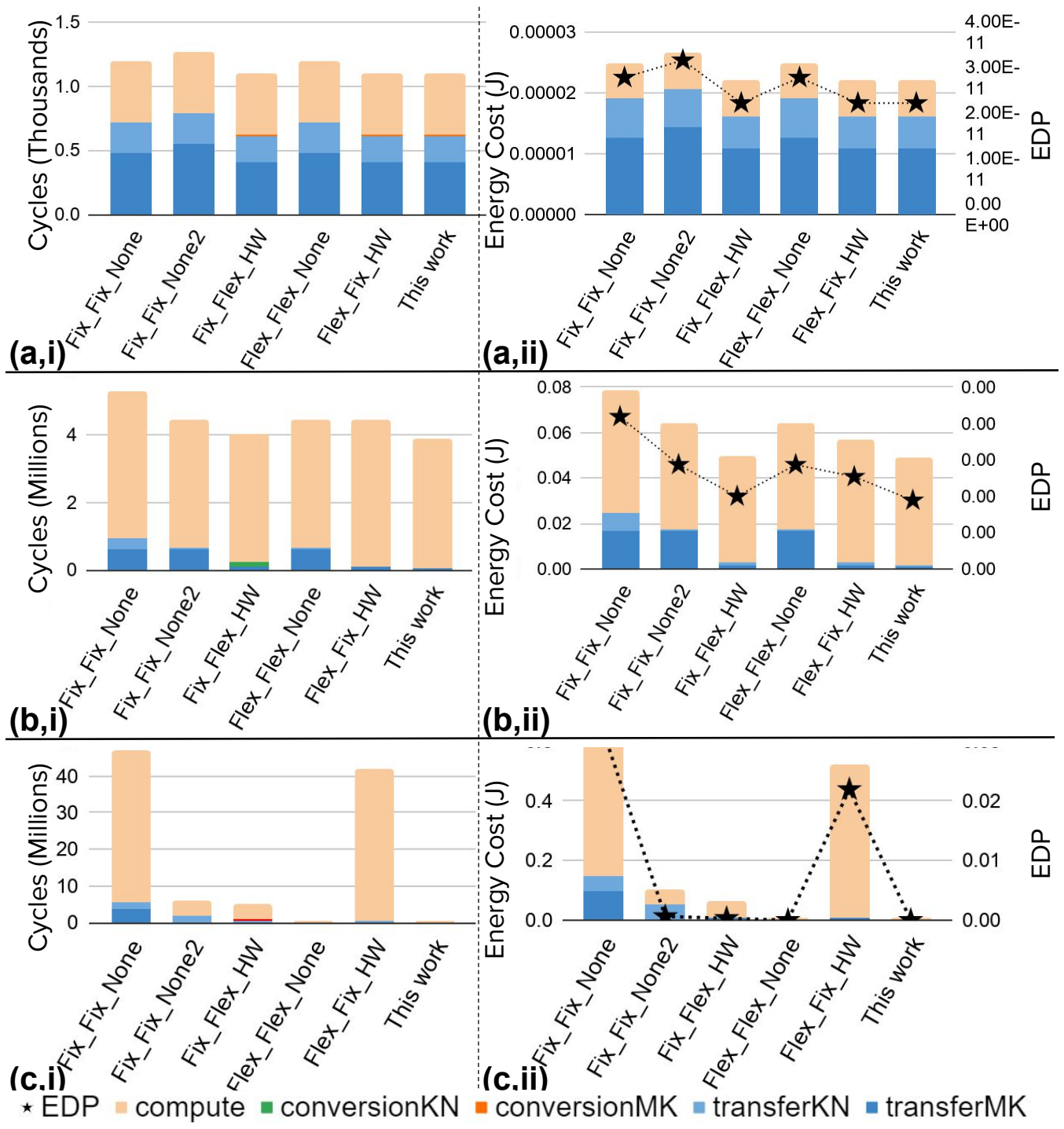

Fig. 12: Breakdown of (a) journals (b) speech2, and (c) m3plates. Part-i shows the performance. Part-ii shows the energy and EDP.

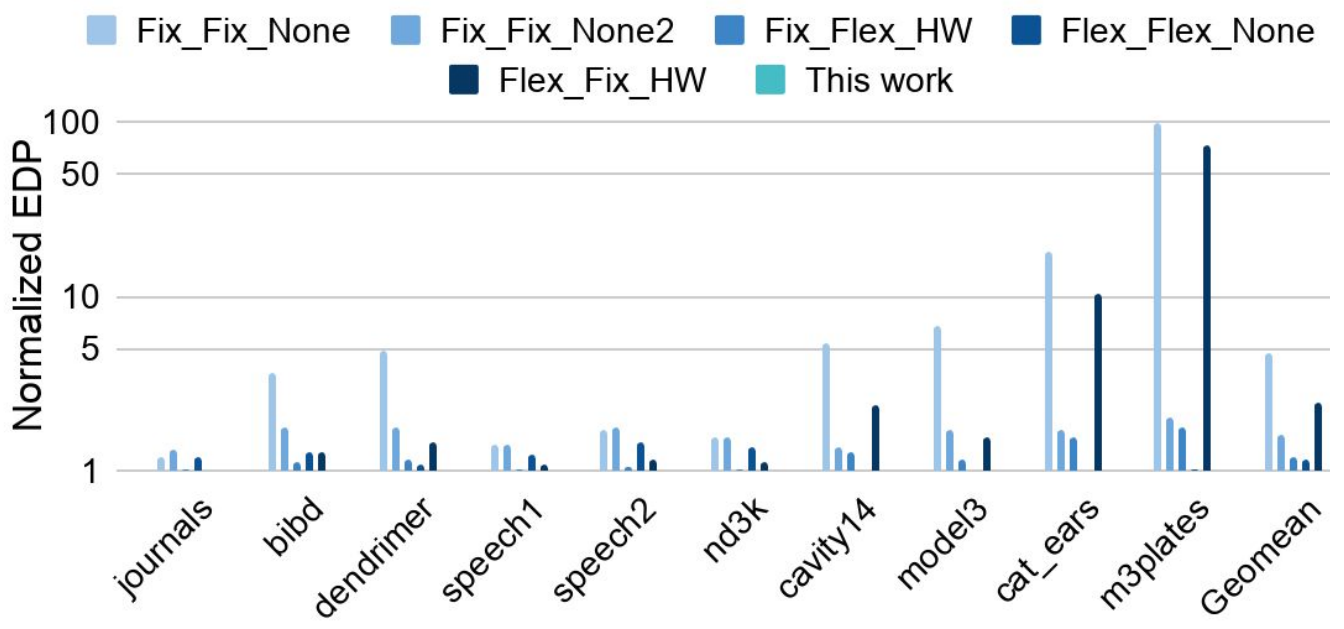

Fig. 13: SpGEMM and SpMM normalized EDP against this work.

Fig. 13 shows the averaged SpGEMM and SpMM normalized EDP of various accelerators against our work. Our work shows geomean reductions of 369%, 63%, 20%, 15%, and 143% over Fix_Fix_None, Fix_Fix_None2, Fix_Flex_HW, Flex_Flex_None, and Flex_Fix_HW respectively; averaging to a ∼122% EDP reduction. The maximum EDP reduction across all workloads is 9860%, 99%, 79%, 44%, and 7338% over Fix_Fix_None, Fix_Fix_None2, Fix_Flex_HW, Flex_Flex_None, and Flex_Fix_HW respectively. The average conversion energy cost is 8.75E-05 J (0.023% of total system energy cost) for Fig. 13 workloads that require conversions

| conv layer id | Input channels (C) | Output channels (K) | input activation (H, W) | filter (R, S) | Normal: input act sparsity | Normal: weight sparsity | 50% Prune (layer): input act sparsity | 50% Prune (layer): weight sparsity | 70% Prune (global): input act sparsity | 70% Prune (global): weight sparsity |
|---|---|---|---|---|---|---|---|---|---|---|
| 1 | 3 | 64 | (32,32) | (3,3) | 0% | 0% | 0% | 50.0% | 0% | 45.4% |
| 2 | 64 | 256 | (32,32) | (1,1) | 56.6% | 0% | 55.5% | 50.0% | 55.0% | 74.8% |
| 3 | 128 | 512 | (16,16) | (1,1) | 63.1% | 0% | 59.2% | 50.0% | 60.4% | 63.4% |
| 4 | 128 | 128 | (16,16) | (3,3) | 52.6% | 0% | 52.0% | 50.0% | 52.3% | 35.3% |
| 5 | 1024 | 256 | (8,8) | (1,1) | 60.2% | 0% | 57.0% | 50.0% | 59.8% | 49.9% |
| 6 | 256 | 256 | (8,8) | (3,3) | 59.4% | 0% | 56.5% | 50.0% | 57.0% | 38.3% |
| 7 | 512 | 2048 | (4,4) | (1,1) | 64.0% | 0% | 61.0% | 50.0% | 41.0% | 88.2% |
| 8 | 512 | 512 | (4,4) | (3,3) | 49.2% | 0% | 47.8% | 50.0% | 43.6% | 98.4% |

(a)

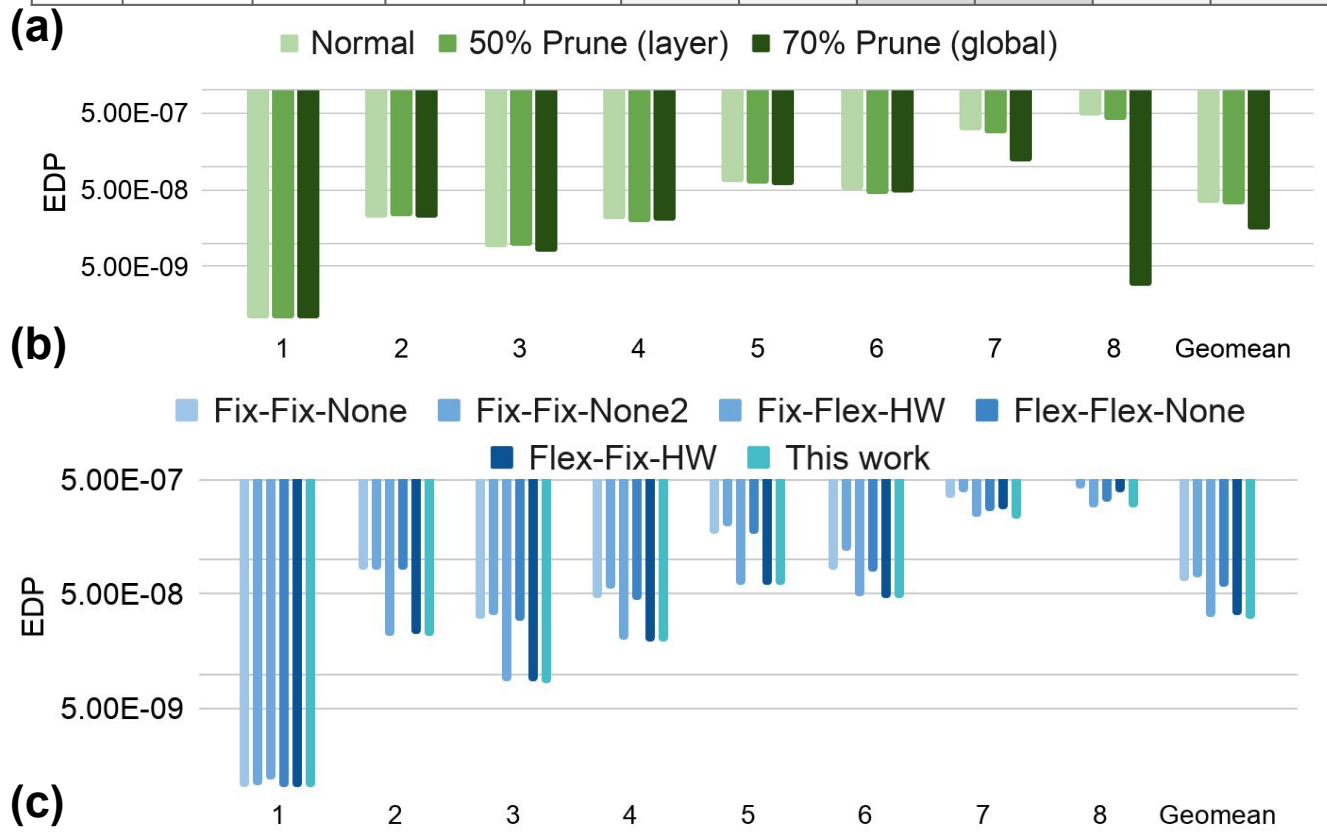

Fig. 14: (a) Convolution layers specifications. (b) EDP of our work with different pruning strategies. (c) Average EDP across the three pruning strategies of our work against different hardware baselines.

with our methodology. Conversion energy cost is negligible because (1) accessing data from DRAM consumes significantly more energy than compute [13], and (2) for matrix multiplications, the conversion complexity is O(MK + KN), while the computation complexity is O(MNK). Part one of Fig. 12 shows system cycles breakdown and part two of Fig. 12 shows system energy cost breakdowns.

## D. Case Study: Convolutional Neural Networks

We train a ResNet50 model using the CIFAR-10 dataset [37]. Fig. 14a shows select convolution layers and their specifications. All layers have a stride of one. To get a wide spectrum of sparsity regions, we apply two different L1 unstructured pruning strategies to induce weight sparsity. The first strategy prunes 50% of values per layer, while the second strategy prunes 70% of values globally; accuracy loss is 0.29% and 0.74% respectively. As shown in Fig. 14a, with global pruning, convolution layers 7 and 8 have significantly higher weight sparsity than the other layers. Input activation sparsity, induced from ReLU, is fairly consistent across different pruning strategies.

For our evaluations, we use a batch size of 64. System configuration is described in Section VII-A. Like TPU, we use im2col to convert convolutions to GEMM operations. Fig. 14b shows the EDP of our proposed work with different pruning strategies. In early layers (1-6), the number of weight elements is relatively small compared to the number of activation elements; hence, the weight sparsity has little impact on the EDP. Layers 7 and 8 have more weight elements than activation elements. Additionally, with global pruning, the two layers are significantly sparser. This allows (1) larger weight pruning impact on EDP by using optimal MCF compression, and (2) better compute utilization, particularly by using the ACF of Dense(**A**)-CSC(**B**) rather than Dense(**A**)-Dense(**B**). This is because the weight matrix (**B**) is much sparser, and will utilize less PE buffer space when stored as CSC (metadata and nonzero data) than Dense (nonzero data and zero-valued data). Fig. 14c shows the average EDP across the three pruning strategies of our work against different accelerator baselines. Since our work enables various MCF (using MINT) and ACF (using accelerator extensions) combinations, we observe on average ∼70% EDP reduction across all baselines.

## VIII. Related Work

**Hardware Accelerators for Sparsity.** Numerous DL accelerators have recently been proposed for sparse computations [14], [15], [18], [20]. There are also tensor algebra accelerators for scientific applications [5], [24], [38]. Recently, NVIDIA's Ampere GPUs added support for limited structured sparsity [25]. The aim of this work is to enable an accelerator to support multiple compression formats; and thus optimize for both memory and compute costs. It can be applied, in principle, over any of the sparse accelerators.

**Format Conversion Support.** The TACO compiler proposes six primitives to concisely represent all tensor formats [12]. Custom hardware modules for (de)compression are common across many applications [9], [39]. However, most custom designs target specific conversions. This limits the flexibility of what the MCFs and ACFs can be. In contrast, MINT proposes to use generic building blocks.

## IX. Conclusion

DL and scientific workloads exhibit vastly different sizes and sparsity. We demonstrate that this leads to different optimal compression formats for storage and compute. Our main contributions include: accelerator extensions for flexible format support, efficient format conversion in hardware using MINT and SAGE predictor to find the optimal format combination. Together, these extensions can enable accelerators deployed in datacenters or HPC systems to run a mix of scientific and DL workloads at high-efficiency.

## X. Acknowledgment

We thank Michael Pellauer, Clay Hughes, Si Hammond, Gokcen Kestor, and Roberto Gioiosa for insightful comments and discussions on this work. We also thank the anonymous reviewers for their valuable feedback. Support for this work was provided through the ARIAA co-design center funded by the U.S. Department of Energy (DOE) Office of Science, Advanced Scientific Computing Research program. Sandia National Laboratories is a multimission laboratory managed and operated by National Technology and Engineering Solutions of Sandia, LLC., a wholly owned subsidiary of Honeywell International, Inc., for the U.S. Department of Energy's National Nuclear Security Administration under contract DE-NA-0003525.